\documentclass[12pt]{iopart}
 \usepackage{graphicx}

\usepackage{iopams}  
\usepackage{xcolor}
\usepackage[normalem]{ulem}
\usepackage{epsfig}
\usepackage{amssymb}

\begin{document}

\title{Exact collective occupancies of the Moshinsky model in two-dimensional geometry}

\author{Arkadiusz Kuro\'s$^1$, Adam Pieprzycki$^2$, Edyta Gawin$^2$, Przemys\l aw Ko\'scik$^2$}

\address{$^1$Institute of Physics, Jan Kochanowski University, ul. Uniwersytecka 7, 25-406 Kielce, Poland \\
$^2$Department of Computer Sciences, University of Applied Sciences, Mickiewicza 8, 33-100 Tarn\'{o}w, Poland}
\ead{arkadiusz.kuros@ujk.edu.pl}

\date{}
\begin{abstract}
In this paper, we investigate the ground state of $N$ bosonic atoms confined in a two-dimensional isotropic harmonic trap, where the atoms interact via a harmonic potential. We derive an exact diagonal representation of the first-order reduced density matrix in polar coordinates, in which the angular components of the natural orbitals are eigenstates of the angular momentum operator. Furthermore, we present an exact expression for the collective occupancy of the natural orbitals with angular momentum $l$, quantifying the fraction of particles carrying that angular momentum. The present study explores how the dependence of collective occupancy relies on angular momentum $l$ and the control parameters of the system. Building on these findings, we examine boson fragmentation into components with different $l$ and reveal a unique feature of the system: the natural orbitals contributing to the correlations are uniformly distributed across all significant $l$ components.

\end{abstract}

\maketitle
 
 \section{Introduction} Following the advent of quantum mechanics, the study of interacting particle systems emerged as a topic of significant interest in the scientific community. In the early years, the focus was on studying natural atom systems with a small number of particles, relying mainly on approximate methods such as Hartree-Fock and variational methods based on trial wavefunctions. With the experimental realization of particle systems trapped in external potentials, the possibility of simulating quantum systems in a controlled manner has emerged. This has resulted in an explosion of interest in theoretical studies of the properties of such systems. Quantum correlation between particles and entanglement have been an important part of this research because of their relevance in quantum engineering and quantum information science. Entanglement is the key to quantum computers, which enable them to solve specific problems much faster than standard computers. The quantum community has studied the correlation and entanglement characteristics of various composite quantum systems.

A notable example is the system of harmonically trapped $N$-bosons with harmonic interactions, also referred to as the Moshinsky atom or harmonium model \cite{ent0,ent1,ent2,ent3,ent4,ent5,ent6,ent7,ent8}. One advantage of the model is that it is exactly solvable. Furthermore, the eigenvalues and eigenvectors of the ground state reduced density matrix of any order have been obtained in closed analytical form \cite{ex33}. The exact solutions provided by the Moshynsky model are useful for understanding many-body effects and evaluating the accuracy of approximate methods. From a physical perspective, however, the Moshynsky model is not merely an artificial toy system; it can serve as an approximation for more realistic systems. This can be achieved, for example, by approximating interactions using their quadratic forms or by directly fitting potentials \cite{gauss}. The correlation properties of the ground state in the Moshynsky system have been studied from various perspectives. However, most studies have been restricted to one-dimensional geometry; for example, they included the study of the effect of interactions on the fraction of bosons in the condensate \cite{cond}, entanglement \cite{mosh0,mosh1,mosh2}, correlation energy and statistical Kutzelnigg coefficient \cite{mosh1}, and Shannon entropy and mutual information \cite{Shanon}. Over the past few years, various mixtures with square interactions have also attracted attention \cite{mix}.

This research undertakes a comprehensive examination of the harmonium system in the two-dimensional case. We derive an exact closed-form expression for the Schmidt expansion of the one-particle reduced density matrix (1-RDM) in polar coordinates, where the angular parts of the Schmidt orbitals are eigenfunctions of the angular momentum operator with eigenvalues $\hbar l$. Here, the Schmidt orbitals and their coefficients are nothing other than the natural orbitals and occupancies, as the eigenvectors and eigenvalues of the 1-RDM are called, respectively. Building on these findings, we analyze correlations in the momentum space of a single particle. To achieve this, we use the collective occupancy of the natural orbitals with $l$. This is the sum of the corresponding occupancies and represents the fraction of particles occupying the $l$ orbitals. We have derived an exact formula for the collective occupancy in the general case of $N$ particles for any interaction strength. This provides an opportunity to gain insight into the phenomenon of fragmentation of the state of $N$ particles into single-particle states of different $l$ depending on the control parameters.

The results are presented in a structured manner. Section \ref{Sec1} provides a closed-form analytical expression for the Schmidt form of the 1-RDM in polar coordinates and discusses the quantities used in the analysis. In Section \ref{Sec2}, we present a discussion of the results and a summary of the main conclusions. Finally, section \ref{Sec3} summarizes the present work.
 

\section{Two-dimensional isotropic Moshinsky model}\label{Sec1} 

\subsection{The $N$-bosons Hamiltonian}
We consider a system of $N$ bosons in a two-dimensional isotropic harmonic potential, interacting with a force proportional to the square of the distance between the particles. The Hamiltonian of this system has the form \begin{equation}\label{Hamiltonian_total}
 {\cal H} = \sum_{i=1}^N\left[-{\hbar^2\over 2m}\nabla^{2}_{\mathbf{r}_{i}} + {m \omega^2\mathbf{r}^2_{i}\over 2}+\sum_{j=i+1}\Lambda|\mathbf{r}_i-\mathbf{r}_j|^2\right],
 \end{equation}
which can be transformed into a dimensionless form using the spatial coordinates expressed in $\sqrt{\hbar/m\omega}$, the interaction parameter $\Lambda$ in $m \omega^2$ and the energy in $\hbar \omega$. 

\subsection{One-particle reduced density matrix}
The integral representation of the one-particle reduced density matrix is as follows
 \begin{equation}
 \rho(\mathbf{r},\mathbf{r}^{'})=\int \Psi^{*}(\mathbf{r},\mathbf{r}_{2},...,\mathbf{r}_{N})\Psi(\mathbf{r}^{'},\mathbf{r}_{2},...,\mathbf{r}_{N})d\mathbf{r}_{2}...d\mathbf{r}_{N}.
 \end{equation}
Their eigenvectors $u_{k}(\mathbf{r})$ (natural orbitals) and eigenvalues $\lambda_{k}$ (occupancies) are determined by the following integral eigenproblem 
 \begin{equation} \int \rho(\mathbf{r},\mathbf{r}^{'})u_{k}(\mathbf{r}^{'})d\mathbf{r}^{'}=\lambda_{k}u_{k}(\mathbf{r}).\end{equation}
These eigenvalues can be interpreted as expansion coefficients in the Schmidt decomposition, while the corresponding eigenfunctions represent the natural orbitals of the system
 \begin{equation}\rho(\mathbf{r},\mathbf{r}^{'})=\sum_{k}\lambda_{k}u^{*}_{k}(\mathbf{r}^{'})u_{k}(\mathbf{r}).\end{equation} The eigenvalues $\lambda_{k}$ are widely employed to quantify correlation effects in many-body systems and to distinguish between various quantum phases, including condensed and fragmented states. In the system under consideration, the one-particle reduced density matrix can be derived analytically in Cartesian coordinates \cite{0} and is conveniently expressed in polar coordinates as follows
\begin{equation}\label{rdm}
 \rho(\mathbf{r},\mathbf{r}^{'})=A \exp \left(-{B\over 2}(r^2+{r^{'}}^{2})+{C\over 2} r r^{'}\mathrm{cos}(\varphi-\varphi^{'})\right),
\end{equation}
with the normalization condition $2\pi\int  r\rho(\mathbf{r},\mathbf{r})dr=1$, where 
\begin{equation} A={\omega\over \pi\gamma}, \ \ \ \ \ \ \ \ B={\omega\over \gamma}+{C\over 2}, \ \ \ \ \ \ \ \ C=\left({1-\omega\over N}\right)^2{(N-1)\over \gamma},\end{equation} 
and 
\begin{equation}\gamma={(N-1+\omega)\over N}, \ \ \ \ \ \ \ \ \omega=\sqrt{1+2\Lambda N}.\end{equation}

\subsection{Diagonal representation of the one-particle reduced density matrix}
The objective of this study is to derive the Schmidt decomposition of Eq. (\ref{rdm}) in polar coordinates, providing a diagonal representation of the one-particle reduced density matrix. To this end, we first expand Eq. (\ref{rdm}) in a Fourier-Lagrange series.
\begin{equation}\label{fl}
 \rho(\mathbf{r},\mathbf{r}^{'})={\rho_{0}(r, r^{'})\over {2\pi}}+\sum_{l=1}{ \rho_{l}(r,r^{'}) \mathrm{cos}[l(\varphi-\varphi^{'})]\over{\pi}},
 \end{equation}
 where the $l$-partial wave component is given by the following integral 
 \begin{equation}\label{fur}
 \rho_{l}(r, r^{'})=\int_{0}^{2\pi} \rho(\mathbf{r},\mathbf{r}^{'}) \mathrm{cos}(l\theta) d\theta,\end{equation}
wherein $\theta=\varphi-\varphi^{'}$, which results in 
 \begin{equation}\label{fur1}
 \rho_{l}(r, r^{'})=2A\pi \exp\left(-{B\over 2}(r^2+{r^{'}}^2)\right)I_{l}\left({C rr^{'}\over 2}\right),\end{equation}
 where we have employed an integral formula for the modified Bessel function of the first kind $I_{l}(z)$, that is \cite{abramowicz}   
 \begin{equation}\int_{0}^{2\pi}d\theta \exp \left(z \cos(\theta) \right) \cos(l \theta )=2\pi I_{l}(z). \end{equation}
A detailed analysis indicates that the Schmidt decomposition of the $l$-partial component, denoted as $\rho_{l}$ in Eq. (\ref{fur1}), can be derived by direct comparison with the Hardy-Hille formula \cite{Hardy}
 \begin{equation}
 \exp \left(-{\left({1\over 2}+{t\over 1-t} \right)(x+y)} \right) I_{\alpha}\left({2 \sqrt{ xyt}\over 1-t}\right)=  \label{hil} \end{equation} $$ = \sum_{n=0} {n! t^{n+\frac{\alpha}{2}}(1-t)\over (n+\alpha)!}(xy)^{\frac{\alpha}{2}}\exp \left(-{(x+y)\over 2} \right) L_{n}^\alpha(x)L_{n}^{\alpha}(y),$$
where $L_{n}^{\alpha}(x)$ is the generalized Laguerre polynonomial and
  \begin{equation}t=-1+{4B\over C^2}\left ( 2B-\sqrt{4B^2-C^2}\right). \end{equation}
By substituting $x = z^2r^2$ and $y = z^2{r^{'}}^2$ into the formula (\ref{hil}), where 
\begin{equation}z={(4B^2-C^2)^{1\over 4}\over \sqrt{2}},\end{equation}
 we derive the matching condition 
\begin{equation}B=\left(1 + 2t(1 - t)^{-1}\right)z^2,  \ \ \  \ \ \ \ \ \ \ \ C=4\sqrt{t}(1 - t)^{-1}z^2,\end{equation} 
which, together with the requirement that Schmidt orbitals be normalized, yields
 \begin{equation}\rho_{l}(r,r^{'})=\sum_{n} \lambda_{nl}v_{nl}(r^{'})  v_{nl}(r),\end{equation}
 with 
 \begin{equation} v_{nl}(r)=\sqrt{{2 n! z^2 \over (n+|l|)!}}(z r)^{|l|}e^{-z^2r^2/2}L_{n}^{|l|}(z^2 r^2),
 \end{equation}
 and \begin{equation}\label{ocup}\lambda_{nl}={A\pi t^{n+|l|/2}(1-t)\over z^{2}}.\end{equation} 
Using the above expressions and the identity $\cos(l\theta) = (\exp(i l\theta) + \exp(-i l\theta))/2$, where $i$ is an imaginary unit, we can represent the 1-RDM as follows
 \begin{equation}\label{SD}
 \rho(\mathbf{r},\mathbf{r}^{'})=\sum_{n=0}^{\infty} \sum_{l=-\infty}^{\infty} \lambda_{nl}u^{*}_{nl}(\mathbf{r}^{'})u_{nl}(\mathbf{r}),
 \end{equation}
 where  \begin{equation}\label{Scm}
 u^{}_{nl}(\mathbf{r})={v_{nl}(r)}{e^{i l\varphi}\over\sqrt{2\pi}}. \end{equation}
 The orbitals (\ref{Scm}) form an orthonormal basis set  
 \begin{equation}\int_{0}^{2\pi} \int_{0}^{\infty}d\varphi dr [r u^{*}_{nl}(\mathbf{r})u^{}_{n^{'}l^{'}}(\mathbf{r})]=\delta_{nn^{'}}\delta_{ll^{'}},\end{equation} and they can be recognized as the natural orbitals of the occupancies (\ref{ocup}). It is worth mentioning that the Schmidt decomposition (\ref{SD}) is not unique due to the double degeneracies appearing in the spectrum of the occupancies $\lambda_{nl} = \lambda_{n|l|}$. Alternative forms of Schmidt orbitals can be constructed for the degenerate points using the method described in \cite{ghi}.

\subsection{Collective occupancies, participation, and correlation}
There are many different quantities for studying the properties of many-body states. In the case of isotropic symmetry, one of them is the collective occupancy, defined as the sum of occupancies with angular momentum $l$ \cite{pk88}
 \begin{equation}
 \eta_l= \sum_n \lambda_{nl}=\frac{n_{l}}{N},
 \label{ocu} \end{equation}
where $n_{l}$ denotes the number of particles with that angular momentum. This quantity determines a fraction of particles with a given $l$. For the present system, the collective occupancy is given by the geometric series ($0 < t < 1$) and can thus be accurately determined in closed analytical form
 \begin{equation} 
 \eta_l=  \frac{ \pi A t^{\frac{|l|}{2}}}{z^2},
 \end{equation}
which gives us a unique opportunity to examine its behavior in the full interaction regime and the general case of $N$ particles. Another closely related quantity is participation, defined as
 \begin{equation} K_{\eta}= \left (\sum_{l} \eta_l^2 \right)^{-1},\end{equation}
which measures the effective number of collective occupancies, i.e., the number of $l$ fragments that contribute significantly. The collective occupancies and this tool offer insight into correlations in the single-particle angular momentum domain. For the same reason as for collective occupancy, this measure can be obtained analytically.
 \begin{equation}\label{par}K_{\eta}=\frac{z^4(1-t)}{\pi^2 A^2 (1+t)}.\end{equation}
 By contrast, participation
 \begin{equation} K=\left ( \sum_{nl} \lambda_{nl}^2 \right)^{-1} =\frac{z^4}{\pi^2 A^2}.
 \end{equation}
It quantifies the effective number of natural orbitals and serves as an indicator of the degree of single-particle correlations \cite{part}. More specifically, it assesses how a subsystem containing one particle correlates with a subsystem composed of the remaining particles.
 
\begin{figure}[h!]
\includegraphics[width=1\columnwidth]{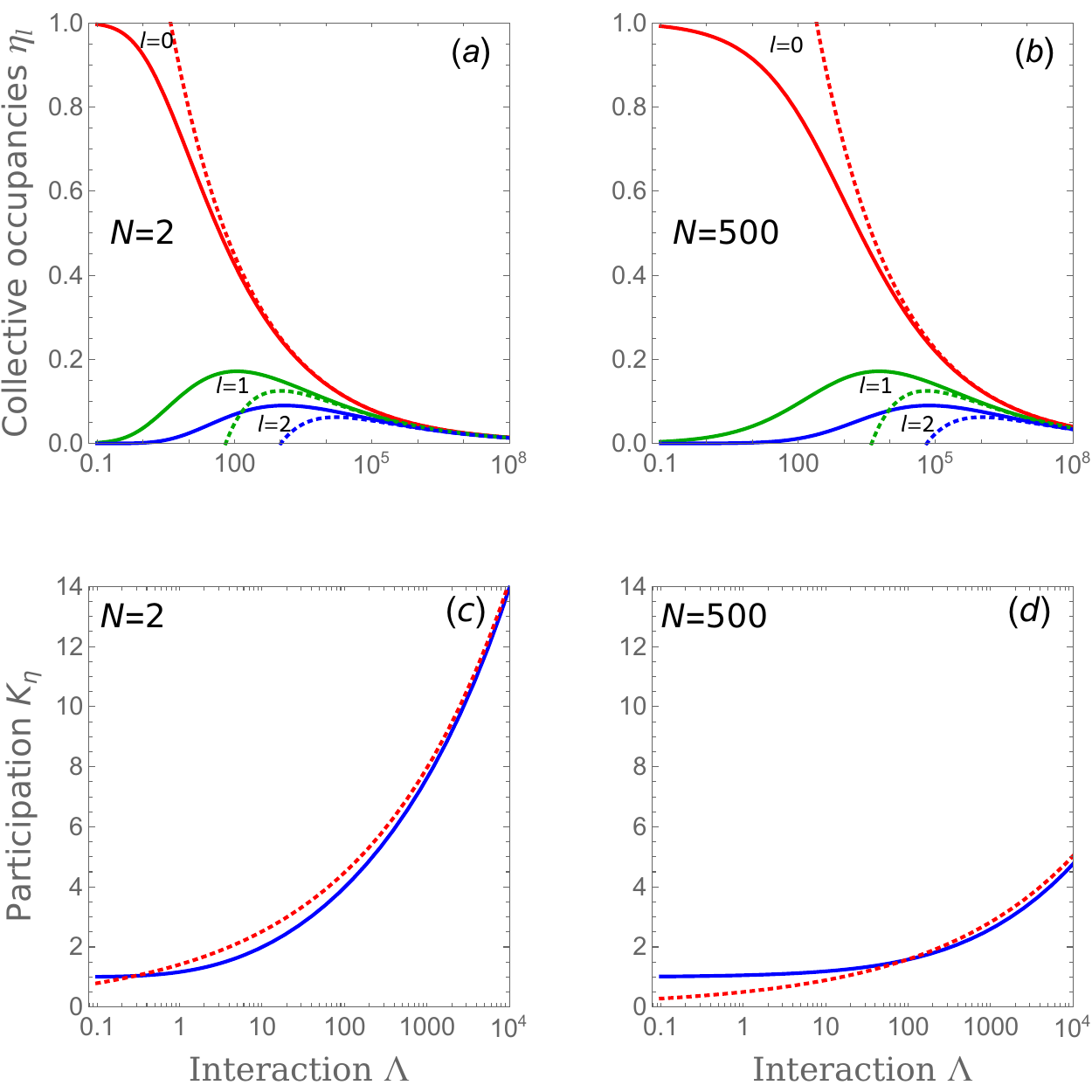}
\caption{Graphs \textbf{(a)} and \textbf{(b)} show the behaviors of collective occupancies as functions of interaction strength $\Lambda$ for two different particle numbers $N=2$ and $N=500$, respectively. The dashed lines represent the approximation results (\ref{as}). The strength of the interaction $\Lambda$ is expressed in units of $m \omega^2$. The graphs \textbf{(c)} and \textbf{(d)} illustrate the corresponding results for the participation $K_{\eta}$, together with its approximation obtained from the expansion to infinity, $\Lambda \to \infty$: $K_{\eta} \approx 2\beta^{-1}(N)\Lambda^{1/4}$ (dashed lines).}\label{Fig1}
\end{figure}
 
 \section{Discusions}\label{Sec2}
\begin{figure}[]
\includegraphics[width=1\columnwidth]{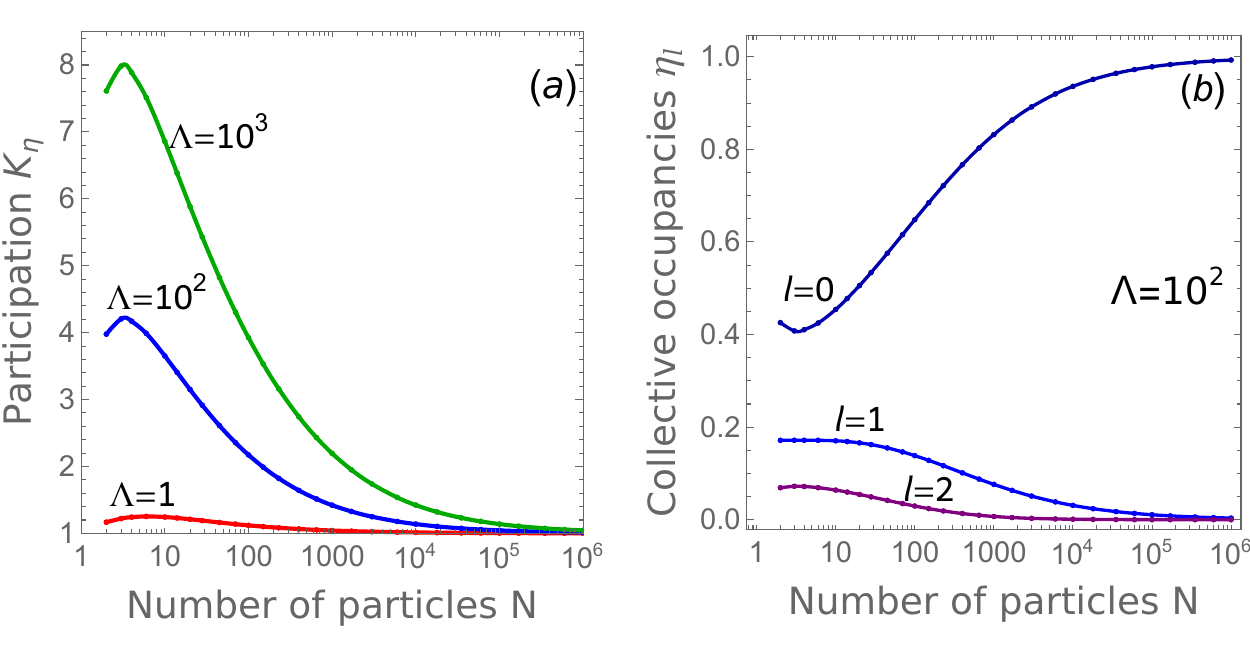}
\caption{Graph \textbf{(a)} shows the participation $K_{\eta}$ obtained for different values of $\Lambda$ as a function of $N$. The plot \textbf{(b)} shows the corresponding behaviors of the three lowest collective occupancies for $\Lambda=10^2$. }\label{Fig2}
\end{figure}
\begin{figure}[h]
 \includegraphics[width=1\columnwidth]{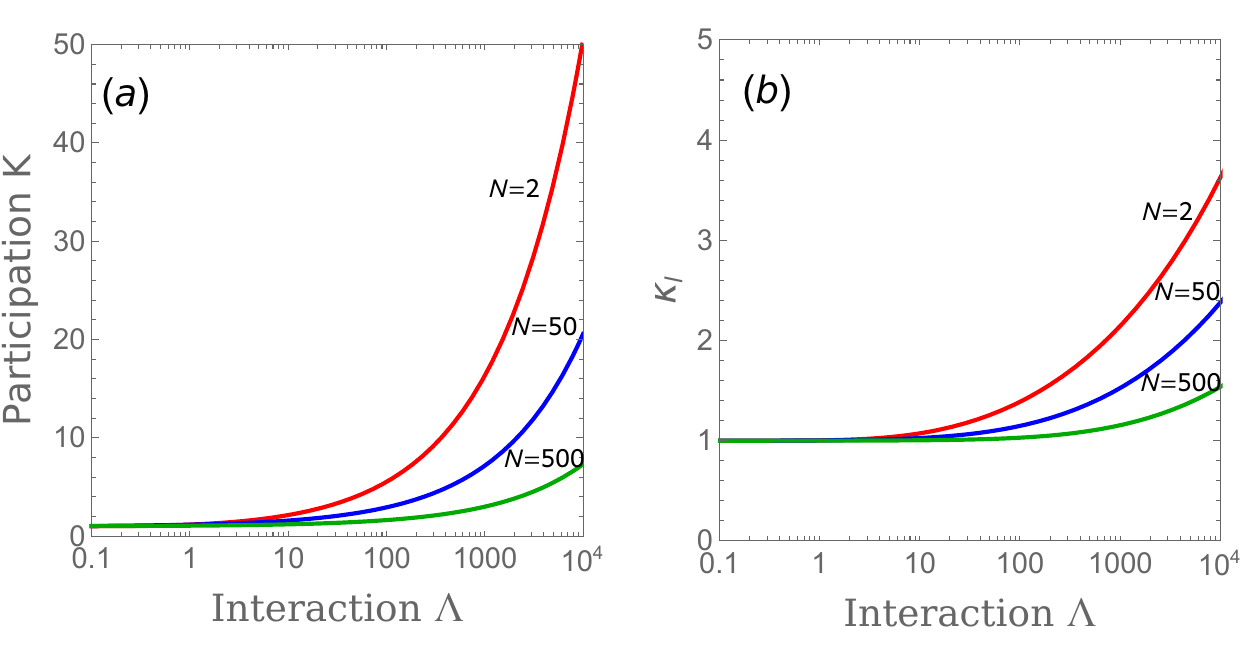}
 \caption{Graphs \textbf{(a)} and \textbf{(b)} illustrate participation ratios $K$ and $\kappa_{l}$, respectively, for particle numbers $N=2, 50$ and $N=500$ as functions of $\Lambda$. The strength of the interaction $\Lambda$ is expressed in units of $m \omega^2$. }\label{Fig3}
 \end{figure}

Figure \ref{Fig1} summarizes the results obtained as a function of the interaction strength $\Lambda$. The upper panel shows the behavior of the collective occupancies for two different numbers of particles. In contrast, the lower panel shows the corresponding results obtained for the participation in Eq. (\ref{par}). Regardless of the number of particles, a similar trend is observed; namely, as the collective occupancy with the lowest angular momentum decreases, the significance of collective occupancies with higher angular momenta increases with $\Lambda$. This means that the number of natural orbitals with higher $l$ needed to capture most of the correlations also increases, which is also manifested by the participation showing an increasing behavior with $\Lambda$. Interestingly, a clustering of collective occupancies is visible. To better understand this phenomenon, we have expanded the collective occupancy $\eta_{l}$ to infinity. The resulting series expansion of $\eta_{l}$ truncated to the second order gives an approximate expression for the behavior of $\eta_{l}$ for large $\Lambda$.
\begin{equation}\label{as}\eta_{l}\approx {\beta(N)\Lambda^{-{1\over 4}}}-{2|l|\beta^2(N) \Lambda^{-{1\over 2}}},\end{equation}
where
\begin{equation}\beta(N)={N^{3/4}\over 2^{1/4}\sqrt{N-1}}.\end{equation}
The results of Eq. (\ref{as}), shown in the top panel of Figure \ref{Fig1} with dashed lines, agree well with the exact results for extremely large $\Lambda$. The dependence on $l$ vanishes gradually with increasing $\Lambda$, and $\eta_{l}$ behaves asymptotically as $\eta_{l} \sim \beta(N)\Lambda^{-1/4}$. By contrast, as shown in the bottom panel of Fig. \ref{Fig1}, the first-order expansion of $K_{\eta}$ to infinity (see the caption to this Figure) provides a reasonable approximation over a wide range of values of $\Lambda$. We can conclude that the number of pieces with different $l$ into which the state of the $N$ bosons is fragmented grows with $\Lambda$ at a rate proportional to $\Lambda^{1/4}$. To better understand the effect of the number of particles, we also study the dependence of the considered quantities on $N$. In Figure 2, our results are presented: the graph \textbf{(a)} shows the behavior of the participation ratio $K_{\eta}$ obtained for various values of $\Lambda$ as a function of $N$, and the graph \textbf{(b)} depicts the collective occupancies obtained for a specific value of $\Lambda$ as a function of $N$. We see that a maximum appears in the behavior of $K_{\eta}$, and the point at which it occurs depends only slightly on $\Lambda$. This point coincides with the minimum of the lowest collective occupancy. Once this local extreme is crossed, the lowest collective occupancy increases, indicating that more and more particles occupy single-particle states with zero angular momentum. In fact, in the limit of the macroscopic number of particles, the lowest occupancy number is $\lambda_{00} \approx 1 - \sqrt{\Lambda/2N}$, which means that only it survives in this limit and, as a result, condensation takes place, $\lambda_{00} \approx 1$. The above approximate formula indicates that as $\Lambda$ increases, the number $N$ at which the system condenses also increases. Surprisingly, we found that the effective number of natural orbitals in the $l$-fragment is independent of $l$, as indicated by the participation \begin{equation} \kappa_{l}=\left(\sum_{n}\left(\frac{\lambda_{nl}}{\eta_{l}}\right)^2\right)^{-1}={1+t\over 1-t},\end{equation}
where $\sum_{n}\lambda_{nl}/\eta_{l}=1$. In addition, our inspection revealed the following 
\begin{equation}\label{un}K=\kappa_{l}K_{\eta},\end{equation}
This shows that the natural orbitals contributing to the correlations originate equally from all relevant $l$ fragments. Based on the definitions of participations utilized, it is easy to verify that Eq. (\ref{un}) does not generally apply. We can assume that this equation is a unique feature of the Moshynky system.
For completeness, we have illustrated in Figure \ref{Fig3} the participations $K$ and $\kappa_{l}$, where data for the same number of particles as in Figure \ref{Fig1} is also included for comparison. The results show that the influence of particle number is most pronounced in cases of strong interaction.
 
 \section{Summary}\label{Sec3}
For a two-dimensional isotropic Moshinsky system, we have obtained the diagonal form of the one-particle reduced density matrix in polar coordinates, where the angular parts of natural orbitals are eigenstates of the angular momentum operator with eigenvalues $\hbar l$. We have analytically derived the formula for the collective occupancy of the natural orbitals with $l$. Using this result, we studied the phenomenon of boson fragmentation into components of different $l$. We revealed a unique feature of the Moshinsky system: the natural orbitals contributing to the correlations are derived equally from all significant $l$ components. Fragmentation with multiple macroscopically occupied $l$ levels is observed in the strong interaction limit. We found that the effect of particle number is particularly strong in this limit.

Since certain interactions, such as finite-size Gaussian interactions, can be modeled using quadratic forms, our findings may provide valuable information on correlation and fragmentation in many-body systems. To gain a deeper understanding, it is crucial to study how the confinement potential, type of interaction, and number of particles affect the fragmentation discussed in this work. Due to the lack of studies along this line, our work may encourage others to investigate this issue.
 

\end{document}